\documentstyle[aps,preprint,epsf]{revtex} \draft
\tightenlines

\makeatletter

\providecommand{\LyX}{L\kern-.1667em\lower.25em\hbox{Y}\kern-.125emX\@}
\let\SF@@footnote\footnote
\def\footnote{\ifx\protect\@typeset@protect
    \expandafter\SF@@footnote
  \else
    \expandafter\SF@gobble@opt
  \fi
}
\expandafter\def\csname SF@gobble@opt \endcsname{\@ifnextchar[%]
  \SF@gobble@twobracket
  \@gobble
}
\edef\SF@gobble@opt{\noexpand\protect
  \expandafter\noexpand\csname SF@gobble@opt \endcsname}
\def\SF@gobble@twobracket[#1]#2{}

\makeatother
\begin{document}

\newcommand{\qs}{q^{*}}

\newcommand{\apeq}[2]{\cong _{#2 }^{#1 }}

\newcommand{\rat}{{f^*_{Qq}}/{f_{Qq}}}

\def\et{{\it et al.}}

\title{Lattice results for the decay constant of heavy-light
vector mesons} 

\author{ Claude~Bernard and Peter Williams}
\address{
Department of Physics, Washington University, St.~Louis, MO 63130, USA
}
\author{ Saumen Datta and Steven~Gottlieb}
\address{
Department of Physics, Indiana University, Bloomington, IN 47405, USA
}
\author{ Carleton~DeTar }
\address{
Physics Department, University of Utah, Salt Lake City, UT 84114, USA
}
\author{ Urs~M.~Heller }
\address{
CSIT, Florida State University, Tallahassee, FL 32306-4120, USA
}
\author{ Craig McNeile }
\address{Dept.\  of Math.\  Sci., University of Liverpool, Liverpool, L69 3BX, 
UK}
\author{Kostas Orginos }
\address{RIKEN-BNL Research
Center,
Brookhaven National Laboratory, Upton, NY 11973-5000} 
\author{ Robert~Sugar }
\address{
Department of Physics, University of California, Santa Barbara, CA 93106, USA
}
\author{Doug~Toussaint }
\address{
Department of Physics, University of Arizona, Tucson, AZ 85721, USA}

\date{\today}

\maketitle
\begin{abstract}
We compute the leptonic decay constants of heavy-light vector mesons
in the quenched approximation.
The reliability of lattice computations for heavy quarks is checked by
comparing the ratio of vector to pseudoscalar decay
constant with the prediction of Heavy Quark Effective Theory in the
limit of infinitely heavy quark mass. Good agreement is found.
We then calculate the decay constant ratio for B mesons:
${f_{B^{*}}}/{f_{B}}= 1.01(0.01)(^{+0.04}_{-0.01})$.
We also
quote quenched $f_{B^{*}}=177(6)(17)$ MeV.
\end{abstract}
\vfill\eject

The symmetries of Heavy Quark Effective Theory (HQET) \cite{0}
show how Quantum Chromodynamics (QCD) simplifies in the limit of infinite
quark mass. For a mesonic system such as the neutral $B$, consisting
of a heavy, but finite mass, anti-$b$ quark and a light $d$ quark, HQET
can be applied with the inverse $b$ mass as a small perturbation parameter.
In particular, the ratio of the decay constants of the \( B^{*} \)
and \( B \) can be calculated. Heavy quark spin symmetry implies
that in the limit of an infinite quark mass, the spins of the quarks
decouple and the vector and pseudoscalar mesons are degenerate, so
the ratio of their decay constants is 1. Perturbative corrections
to this ratio are also calculable within the HQET framework.

In this paper, we study the heavy-light vector and pseudoscalar decay
constants in quenched lattice QCD. Because computational restrictions
limit the range of heavy quark masses that are used in our simulations,
the data must be extrapolated to the $B$ mass (or interpolated between
the heavy-light data and a static-light point). The lattice
calculations also inherently require extrapolations to the continuum
limit of zero lattice spacing. The comparison of the lattice calculation
and the HQET calculation of the ratio of the vector and pseudoscalar
decay constants tests the consistency of the treatment of heavy quarks
in lattice QCD. Calculations of heavy-light vector
decay constants on the lattice have been carried out previously
by several other groups \cite{9}. 

In earlier work \cite{1}, we computed pseudoscalar decay constants
only. Here we extend that analysis to include vector mesons. Since
our aim is to test the consistency of these lattice simulations with
the results of HQET, we confine the analysis to the quenched data sets,
the details of which are summarized in Table \ref{table:lattice details}.
The parameters of the generation of these lattices, gauge fixing,
and quark propagator determination are found in \cite{2,1}. 
We use unimproved Wilson valence quarks. {}``Smeared-local{}'' (SL)
and {}``smeared-smeared{}'' (SS) vector meson propagators are calculated
for the heavy-lights. These propagators are also calculated for the
heavy-light and light-light pseudoscalar mesons. The heavy-light pseudoscalar
decay constant $f_{Qq}$ is defined by \[
\left\langle 0\left| A^{cont}_{0}(0)\right| 
P,\vec{p}=0\right\rangle =-if_{Qq}M_{Qq}\]
where\[
A_{0}^{cont}(0)=\overline{Q}\gamma _{0}\gamma _{5}q\]
 is the 0\( {}^{th} \) component of the of the axial current at point
\textbf{x} = 0, and \( \left| P,\vec{p}=0\right\rangle  \)
is a zero 3-momentum pseudoscalar bound state of heavy quark \( \overline{Q} \)
and light quark
\( q \) with mass \( M_{Qq} \). 
We define the vector decay constant
in exact analogy to the pseudoscalar decay constant (this
is standard in HQET) to simplify the
interpretation of the ratio: \[
\left\langle 0\left| V^{cont}_{i}(0)\right| V,\vec{p}=0,\epsilon 
\right\rangle = \epsilon_i f^{*}_{Qq}M^{*}_{Qq}\]
where \( V^{cont}_{i} \) is a spatial component of the continuum
vector current,\[
V^{cont}_{\mu }=\overline{Q}\gamma _{\mu }q\]
and \( \left| V,\vec{p}=0,\epsilon \right\rangle  \)
is the vector meson state with zero 3-momentum, mass \( M^{*}_{Qq} \),
and polarization $\epsilon$.

The vector propagators have the same relation to \( f_{B^{*}} \)
as the pseudoscalar propagators have to \( f_{B} \). The light-light
pseudoscalars are used to set the scale (through \( f_{\pi }) \)
and the physical value of the hopping parameter of the degenerate
up/down quarks (through \( m_{\pi } \)). 

Each pair of SL and SS propagators for a particular mass combination
is fit simultaneously and covariantly to
single exponential forms sharing the same mass; {\it i.e.}, we
make three-parameter fits. The time ranges used in these fits were
varied to produce different fits (typically 8--10 of them) that provided
reasonable confidence levels for both vector and pseudoscalar decay constants. 
The alternate fit ranges were then used to fit the ratio of the decay
constants as discussed below. A preferred fit range was selected from
the acceptable alternatives by choosing a range that provided a good
blend of high confidence level and small statistical error for the
ratio fit. For each ratio derived from these fits, the standard deviation
of the alternate fit ranges was added in quadrature to the raw statistical
error of the preferred fit to produce a measure of the statistical
uncertainty in the ratio that reflects the different possible plateau
regions. 

To relate the matrix elements measured on the lattice to their continuum
counterparts we use the perturbative renormalization factors for heavy-light
currents calculated by Kuramashi \cite{8}. These renormalization
factors include a dependence on the quark mass, which for large quark
masses produces an approximately 100\%  difference in the one loop coefficients
compared to those in the massless quark limit. We adjust the values
calculated in \cite{8} to correspond to our definition of the mean
link, \( u_{0} \), in terms of the critical hopping parameter, \( 
u_{0}=1/8\kappa _{c} \).
The mass dependence of the renormalization factors is more important
for the individual decay constants, but still provides a meaningful
improvement to the calculation of the decay constant ratios, where
only the ratio of the vector and axial current renormalizations is
relevant. As in \cite{1}, we adjust the measured meson pole mass upwards
by the difference of the heavy quark kinematic mass and the heavy
quark pole mass. This allows us to estimate the kinetic mass of the
meson while only looking at its zero-momentum state.

A new element that is added to the previous analysis of this data
is the choice of \( q^{*} \), the momentum scale that satisfies
\cite{6}: \[
ln(q^{*2})=\frac{\int d^{4}qf(q)ln(q^{2})}{\int d^{4}qf(q)}\]
 where \( f(q) \)  is the integrand of a 1-loop current renormalization.
Evaluating the coupling (we use \( g^2_{V} \) defined in terms
of the plaquette \cite{6,10}) at \( \qs  \) and using that coupling
to evaluate the renormalization should reduce higher order effects.
The values of \( \qs  \) for the heavy-light currents have not been calculated.
In Ref.~\cite{1}, Hernandez and Hill's result \cite{4} 
for the tadpole-improved static-light axial current scale
(\( \qs =2.18/a \), a value
close to the tadpole-improved light-light
axial current scale, \( \qs =2.32/a \)\cite{4p}) was used to argue
that \( \qs  \) was only mildly mass dependent. 
The light-light \( \qs  \)
was then used for the heavy-lights. Hernandez and Hill's calculation
has recently been repeated \cite{3c} (see also \cite{3b}), with a
rather different result, \( \qs \sim 1.4/a \). We believe
this static-light \( \qs  \) is likely to be more appropriate for
the heavy-lights than the light-light \( \qs  \), and we use it here.
The new calculation of the static-light \( q^{*} \) includes the
continuum part of static-light current, which gives rise to an \( am_{Q} \)
dependence. For the axial current this dependence is weak enough that
a constant value of \( q^{*} \) from \cite{3c} (\( \qs \approx 1.43/a \))
can be used reliably, and this has been done in \cite{3a}. However,
the \( am_{Q} \) dependence of \( q^{*} \) for the static-light
vector current is more pronounced, so here the scale was calculated
for each heavy kappa for both the vector and pseudoscalar case. We
compare the value of \( f_{B} \) obtained from the mass dependent
\( \qs  \) scheme with the results \cite{3a} for the \( \qs \cong 1.43/a \)
scheme\footnote{%
To quote \( f_{B} \) we had to maintain the distinction between lattice
sets C \& CP, since the static points of these lattices are calculated
differently. This is irrelevant for the ratio of the decay constants
computed here, but is necessary for this consistency check.
} as a consistency check.

To help estimate systematic uncertainties we use three different chiral
fits, in which we extrapolate the results at the light quark kappas
used in the simulation to the kappa appropriate to physical light
quarks, as determined by the pion mass. The first of these, from which
the central value for the ratio is taken and which will be referred
to as the standard analysis, uses quadratic fits vs. \( am_{2} \)
(light quark kinematic mass) for \( m^{2}_{\pi } \), and linear fits
vs. \( am_{2} \) for \( f_{\pi } \), \( M_{Qq} \), and \( f_{Qq} \).
The rationale for these
choices is discussed in \cite{1}. The first of the alternate analyses
has quadratic fits for \( m^{2}_{\pi } \) and \( f_{\pi } \) with
all other fits linear, and the second has quadratic fits for \( m^{2}_{\pi } \),
\( f_{\pi } \), and \( f_{Qq} \). The difference of these chiral
fits is used to assess the systematic error in the choice for the
standard analysis. 

For each set of lattices, the ratio of \( f\sqrt{M} \) for the vector
and pseudoscalar mesons at each heavy kappa is calculated. 
For each heavy $\kappa$, we then divide out \( 1-{g^{2}}/{(6\pi ^{2})} \),
the leading order HQET correction to the ratio \cite{5}, 
using \( g_V^{2} \) evaluated on the lattice at the \( \qs  \) 
appropriate to \( m_{B} \). The
resulting data is fit to the three parameter function \begin{equation}
\label{3pfit}
\frac{b+c/M}{1+d/M}
\end{equation}
 Since the decay constants each have a \( {1}/{M} \) expansion
in HQET, this fitting function can be viewed as the ratio of the first
two terms from the individual expansions. If the data produce the
correct static limit, the constant term in the numerator, \( b \),
should be 1. Table \ref{tabe: 3 parameter intercepts} shows the value
of \( b \) for the standard analysis of each set of lattices. Note
that all the results are consistent with 1. The errors are quite large
on the coarsest lattices at \( \beta =5.7 \), but are much smaller on
the finer lattices. 

\begin{figure}
\epsfxsize=0.99 \hsize
\epsffile{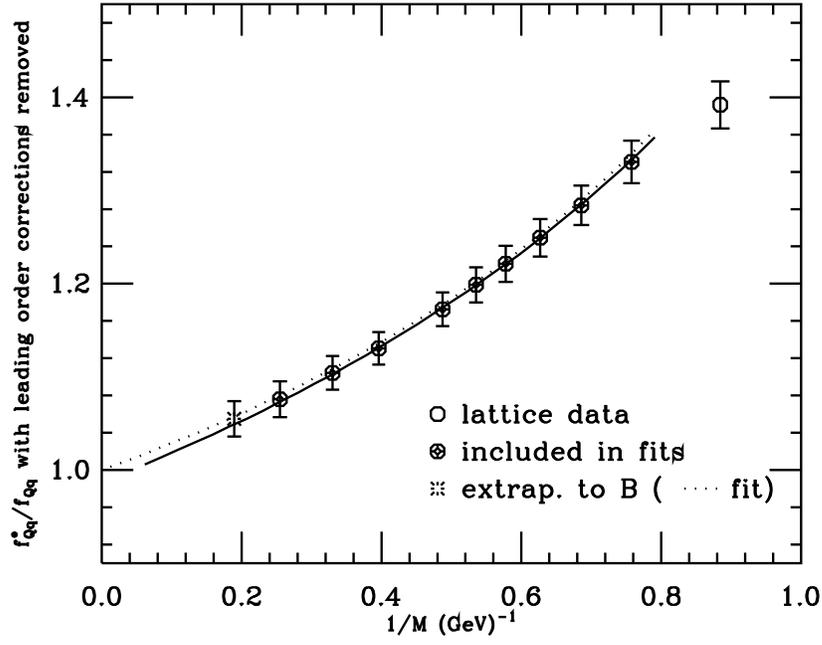}
%{\centering \resizebox*{1\textwidth}{!}{\includegraphics{comp.ps}} \par}
\caption{\label{fig: extrap comparison}Comparison of three-parameter fits and
two-parameter fits for the \protect\( \beta =6.52\protect \) lattices.
Perturbative HQET corrections have been removed. Dotted line enforces the
HQET result that \protect\( f^*_{Qq}/f_{Qq}=1\protect \) at
\protect\( M=\infty \protect \). Solid line allows the \protect\( M=\infty 
\protect \)
value to be free. }
\end{figure}

 From now on we assume consistency with HQET and use the two parameter
fitting function Eq.~(\ref{3pfit}) with $b=1$
to extract the ratio \( {f_{B^{*}}}/{f_{B}} \).
The difference of these fitting methods can be seen in 
Fig.~\ref{fig: extrap comparison}.
The final two-parameter fit is then interpolated to the $B$ mass and the
leading order perturbative correction is reinserted. This result still includes
the ratio of the square roots of the \( B^{*} \) and \( B \)
masses. Removing this gives us a value of the ratio for each set of
lattices, which must then be combined and extrapolated to zero lattice
spacing.

\begin{figure}
\epsfxsize=0.99 \hsize
\epsffile{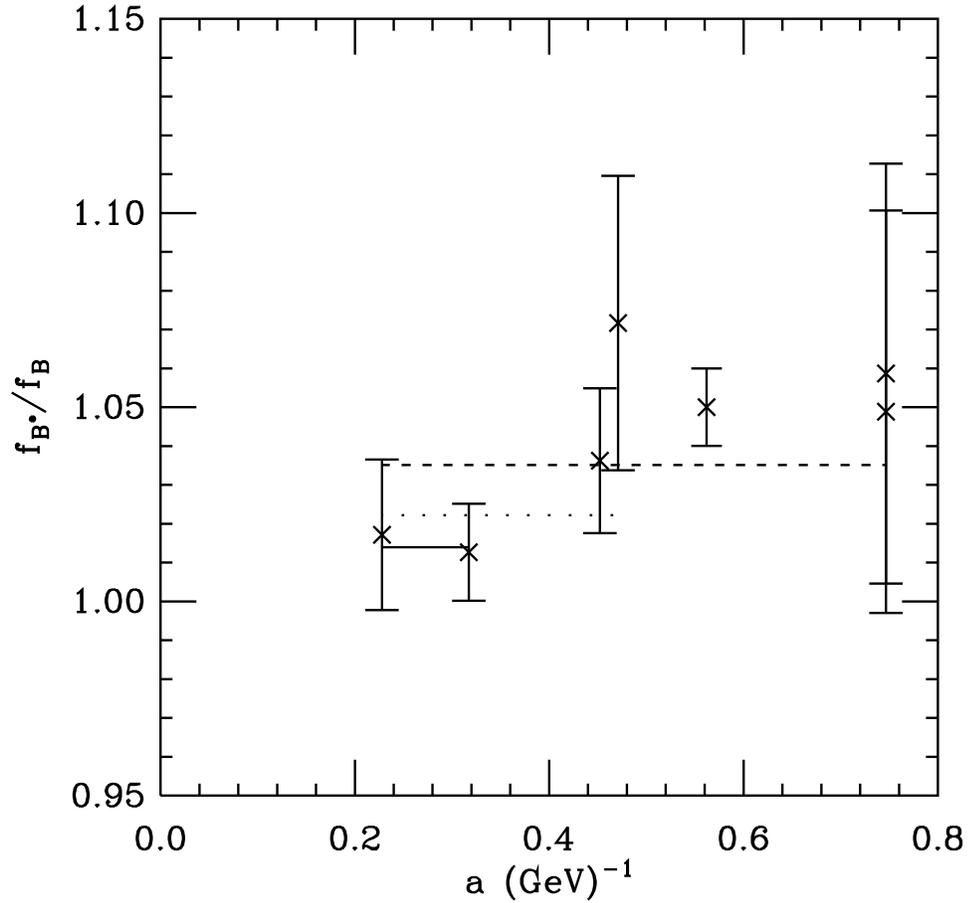}
%{\centering \resizebox*{1\textwidth}{!}{\includegraphics{lat_ext.ps}} \par}
\caption{\label{fig: three plots}Three constant
fits to estimate the lattice spacing
extrapolation error in the ratio. The fit to $a < 0.4$ (solid line) 
is taken as the
central value. The data are for the standard analysis. The fit ranges
are 0.2 to 0.4 \protect\( ({\rm GeV})^{-1}\protect \) (solid line) , 
0.2 to  0.5 \protect\( ({\rm GeV})^{-1}\protect \) (dotted line), and 0.2 
to 0.75 \protect\( ({\rm GeV})^{-1}\protect \)
(dashed line) 
}
\end{figure}

\begin{figure}
\epsfxsize=0.99 \hsize
\epsffile{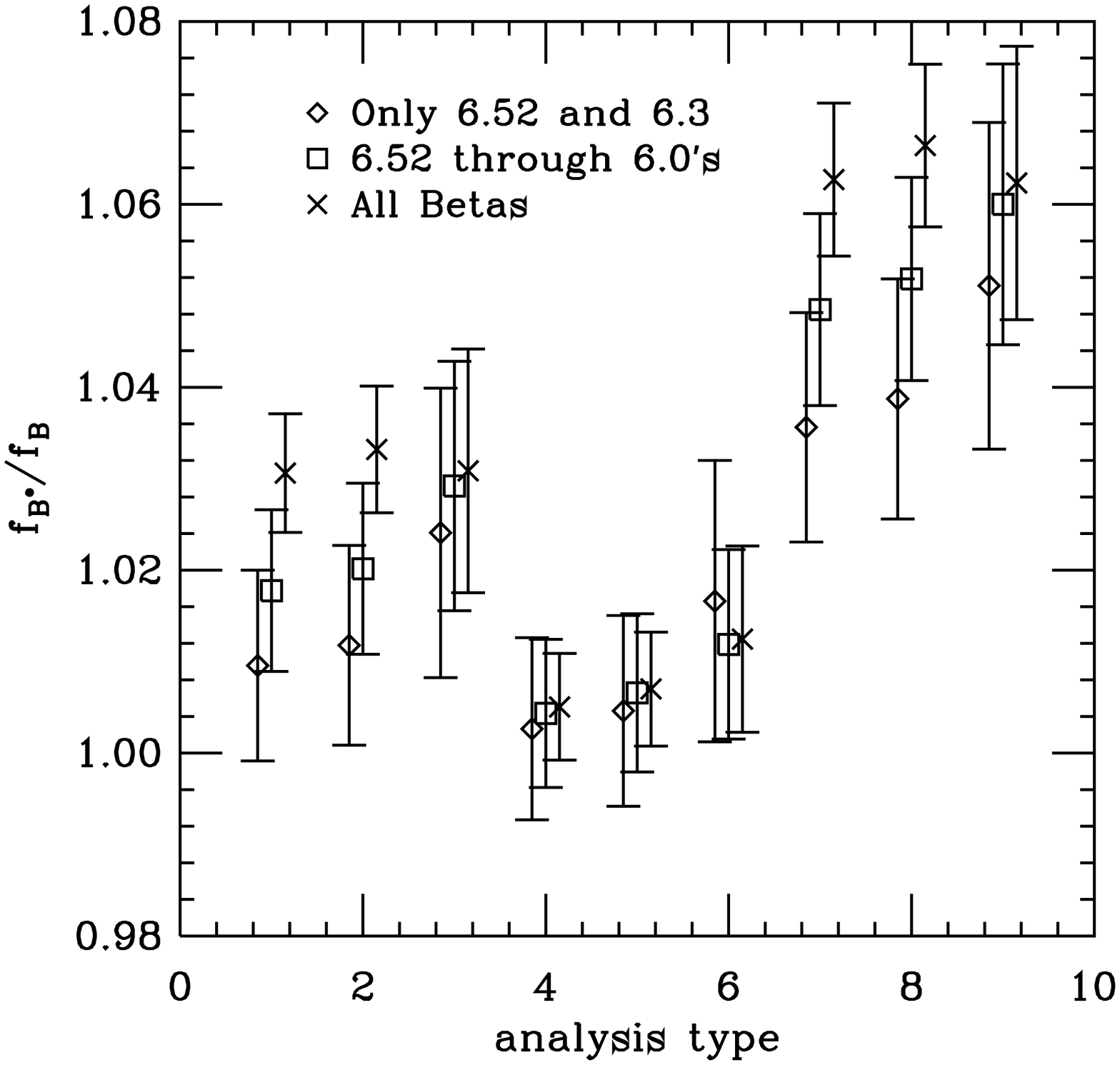}
%{\centering \resizebox*{1\textwidth}{!}{\includegraphics{sim_2.ps}} \par}
\caption{\label{fig: different final plots}Plot of the final data from the
different types of analyses. See Table~\ref{tab: final fit blurbs} for
descriptions of each analysis.}
\end{figure}

The fits shown in Fig.~\ref{fig: three plots} are different possible
lattice spacing extrapolations for the continuum value of the ratio.
These data are the result of the standard analysis on each set of
lattices, but the general features of the plot are generic for all
the analyses, as can be seen in Fig.~\ref{fig: different final plots}.
We use constant fits over different intervals
in the lattice spacing: 0.2 to 0.4, 0.2 to 0.5, and 0.2 to 0.75 
\( ({\rm GeV})^{-1} \).
We do not include a linear fit, as analysis of the new data sets described
in \cite{3a} suggests that the constant fits provide a good measure
of the lattice spacing extrapolation uncertainty. For each set of
analyses, the fit to the interval containing only the values from the
two sets with finest lattice spacing ($0.2 < a < 0.4 ({\rm GeV})^{-1}$) 
is taken to be the central
value for that ratio.

The systematic errors are obtained from various alternative analyses
(see Table \ref{tab: final fit blurbs}). 
The discretization error is
estimated by computing the difference between the average 
of the two finest lattices and the average over all lattice spacings.
The three constant fits for the standard analysis are shown 
in Fig.~\ref{fig: three plots},
and the results of the three fits for each of the alternate analyses
can be seen in Fig.~\ref{fig: different final plots}. We estimate
the lattice spacing extrapolation error as the largest difference
of the three constant fits, which is \( \cong _{-0}^{+0.02} \) for
the standard analysis. 

Higher order perturbative effects are a second source 
of systematic error. This error is estimated by taking the 
difference of the standard
analysis with the analysis performed at different values of \( \qs  \).
In particular, we compare the standard analysis results to the two
alternate analyses 4 and 7 of Table \ref{tab: final fit blurbs},
where \( \qs  \) is adjusted down and up, respectively. The comparison
can be seen in Fig.~\ref{fig: different final plots}. We estimate
the perturbative error to be \( \apeq{+0.03}{-0.01} \). 

The final significant
contribution to the systematic error comes from the chiral extrapolations.
Our estimate of the systematic error involved in this extrapolation
procedure is found by taking the larger difference of the central
value and the two alternate chiral fits described above. This systematic
error can be seen in fig. \ref{fig: different final plots} by comparing
fit number 1 to 3, 4 to 6, or 7 to 9. We estimate the error from chiral
extrapolation in our central value of the ratio as \( \apeq{+0.02}{-0} \).
We do not include an analysis of the other sources of error mentioned
in \cite{1} (difference of magnetic mass and kinetic mass, higher
order lattice extrapolation fits, and finite volume effects) because
they are negligible for the ratio \( {f_{B^{*}}}/{f_{B}} \).

We combine the three sources of systematic error as if they were completely
independent, because we see in Fig.~\ref{fig: different final plots}
that the results of the different changes made in the analysis are
not significantly correlated. This gives us our final quenched value
of \( {f_{B^{*}}}/{f_{B}} \):\[
1.01(0.01)(^{+0.04}_{-0.01})\, .\]
In HQET the leading order (in \( {1}/{M} \)) value of 
the ratio is \( 1-\frac{g^{2}(m_{B})}{6\pi ^{2}}\approx 0.96 \),
using $\Lambda^{(5)}=0.208$ GeV \cite{PDB}.
Note that this is
less than 1 because the perturbative correction is negative. However
the results of our simulations suggest that the ratio for the \( B \)
is more likely to be greater than or equal to 1. Neubert has calculated
that the value of the ratio \( {f_{B^{*}}}/{f_{B}} \)
using the
subleading order terms in the \( {1}/{M} \) expansion to
be \( 1.07\pm 0.03 \) \cite{7}, which is consistent with our result.

Using the same analysis, we find \( f_{B}\cong 175(7) \)MeV, where
statistical error only is shown. This should be compared with the
current MILC value \( f_{B}=173(6)(16) \)MeV \cite{3a}. The latter
includes improved action data and a complete systematic error analysis
and, therefore, should be taken as the most up-to-date MILC value\footnote{%
We note that the central value of \( f_{B} \) in \cite{3a} is considerably
higher than the value \( 157(11)(_{-9}^{+22})(_{-0}^{+21}) \)MeV
quoted in \cite{1}. That difference is due to new data (including
improved action) and new analysis (including update of \( \qs  \))
as explained in \cite{3b} and \cite{3a}. 
} for quenched \( f_{B} \). However, the consistency of the current
analysis with the previous calculation is comforting and indicates,
among other things, that the use of a scheme in which \( \qs  \)
varies with heavy quark mass has no drastic effects. 

We also report a value for \( f_{B^{*}} \). This
quantity was not calculated in \cite{1} or \cite{3a}, so we perform
a more detailed analysis including estimates of systematic errors
analogous to that performed in \cite{3a}.  
This gives \( f_{B^{*}}=177(6)(17) \)MeV
in the quenched approximation.

These results for  $f_{B^{*}}$ and $f_{B^{*}}/f_B$ are
in qualitative agreement with what is expected.
However, the main point of this paper is not the computation of
the quenched decay constants
at the $B$ mass, but the extrapolation of our results to infinite heavy
quark mass.  The agreement of $f^*_{Qq}/f_{Qq}$ in this limit with
the HQET prediction is an indication that the present
treatment of the heavy quark on the lattice is consistent.

Calculations for this project were performed
at ORNL CCS, SDSC, Indiana University,  NCSA,
PSC, MHPCC, CTC, CHPC (Utah),
and Sandia Natl.\ Lab.
This work was supported by the U.S.\ Department of Energy under
grants
DE-FG02-91ER-40628,      %C. Bernard
DE-FG02-91ER-40661,      %S. Gottlieb
DE-FG02-97ER-41022       %U. Heller
and
DE-FG03-95ER-40906,       %D. Toussaint
and by the National Science Foundation under grants
PHY99-70701              %C. DeTar
and
PHY97--22022.            %R. Sugar

\begin{table}

\caption{\label{table:lattice details}Summary of quenched Wilson action 
lattices}

{\centering \begin{tabular}{|c|c|c|c|}
\hline 
set&
\( \beta  \)&
size&
\# confs.\\
\hline
\hline 
A&
5.7&
\( 8^{3}\times 48 \)&
200\\
\hline 
B&
5.7&
\( 16^{3}\times 48 \)&
100\\
\hline 
E&
5.85&
\( 12^{3}\times 48 \)&
100\\
\hline 
C&
6.0&
\( 16^{3}\times 48 \)&
100\\
\hline 
CP&
6.0&
\( 16^{3}\times 48 \)&
305\\
\hline 
D&
6.3&
\( 24^{3}\times 80 \)&
100\\
\hline 
H&
6.52&
\( 32^{3}\times 100 \)&
60\\
\hline
\end{tabular}\par}
\end{table}

\begin{table}
\caption{\label{tabe: 3 parameter intercepts}Agreement of the static limit
of the 3 parameter ratio fits with 1.}
{\centering \begin{tabular}{|c|c|}
\hline 
Set of Lattices (\( \beta ) \)&
Static limit of fitting function (including plateau uncertainties)\\
\hline
\hline 
H(6.52)&
\( 0.99\pm 0.05 \) \\
\hline 
D(6.3)&
\( 0.99\pm 0.04 \) \\
\hline 
CP(6.0)&
\( 1.05\pm 0.04 \)\\
\hline 
C(6.0)&
\( 1.08\pm 0.08 \)\\
\hline 
E(5.85)&
\( 0.89\pm 0.08 \)\\
\hline 
B(5.7/16)&
\( 0.70\pm 0.42 \)\\
\hline 
A(5.7/8)&
\( 1.04\pm 0.25 \)\\
\hline
\end{tabular}\par}

\end{table}

\begin{table}

\caption{\label{tab: final fit blurbs}Explanation of the Analysis types from
Fig~\ref{fig: different final plots}}

{\centering \begin{tabular}{|c|c|}
\hline 
Analysis Number&
Details of Analysis\\
\hline
\hline 
1&
standard analysis\\
\hline 
2&
Change \( f_{\pi } \) fits from linear to quadratic\\
\hline 
3&
Change \( f_{\pi }\) and \( f_{Qq} \)  fits from linear to quadratic\\
\hline 
4--6 &
Same as 1--3 with \( \qs  \) chosen so that its mean value for
the heavy kappas is $1/a$\\
\hline 
7--9 &
Same as 1--3 with \( \qs  \) chosen twice as large as the standard
analysis\\
\hline
\end{tabular}\par}
\end{table}

\begin{thebibliography}{10}
\bibitem{7}M.\ Neubert, Phys.\ Rev.\ {\bf D46}, 1076 (1992). 
(The value we quote for \( \rat  \) at the B mass is numerically the
same as the value of \( {f_{B^{*}}\sqrt{m_{B^{*}}}}/{f_{B}\sqrt{m_{B}}} \)
as calculated in this reference. The addition of the mass dependence
makes no difference to the accuracy we want.)
\end{thebibliography}
\end{document}